\documentstyle[12pt]{article}
\newcommand{\be}{\begin{equation}}
\newcommand{\ee}{\end{equation}}
\newcommand{\bea}{\begin{eqnarray}}
\newcommand{\eea}{\end{eqnarray}}
\newcommand{\nen}{\nonumber \\ \relax}
\newcommand{\nenn}{\nonumber \\[2mm]}
\newcommand{\nennn}{\nonumber \\[3mm]}

\newcommand{\forcepar}{{\hskip 10pt\vskip -15pt}}
\newfont{\headfont}{cmbx10 scaled 1440}
\newfont{\namefont}{cmr10}
\newfont{\initialfont}{cmr10 scaled 1200}
\newfont{\addfont}{cmti7 scaled 1440}
\newfont{\boldmathfont}{cmbx10}
\newfont{\figfont}{cmr7 scaled 1200}

\newcommand{\seq}{\ =\ }
\newcommand{\seqq}{&\seq &}
\newcommand{\qq}{&\qquad &}
\newcommand{\pls}{\ +\ }
\newcommand{\mi}{\ -\ }
\newcommand{\seqv}{\ \equiv\ }
\newcommand{\half}{\frac{1}{2}}
\newcommand{\inv}[1]{\frac{1}{#1}}

\newcommand{\ca}{{\cal A}}
\newcommand{\cb}{{\cal B}}

\newcommand{\cf}{{\cal F}}

\newcommand{\co}{{\cal O}}

\newcommand{\IR}{{\rm {I \kern -0.23em R}}}
\newcommand{\IC}{{I \kern -0.65em C}}

\newcommand{\ap}[1]{{\it Ann. Phys.} {\bf #1}}
\newcommand{\np}[1]{{\it Nucl. Phys.} {\bf B#1}}
\newcommand{\cmp}[1]{{\it Commun. Math. Phys.} {\bf #1}}
\newcommand{\ijmp}[1]{{\it Intl. J. Mod. Phys.} {\bf A#1}}

\newcommand{\cqg}[1]{{\it Class. Quan. Grav.} {\bf #1}}

\newcommand{\pl}[1]{{\it Phys. Lett.} {\bf #1B}}

\newfont{\headfontb}{cmbx10 scaled 1728}
\newcommand{\pint}{{\int\hskip -11.3pt \int}}

\newcommand{\fd}{{\bf d}}

\newcommand{\dual}{\,^\star}
\newcommand{\ym}{{YM}$_2$~}
\newcommand{\cp}{{\cal P}}
\newcommand{\cq}{{\cal Q}}
\newcommand{\tq}{{\tilde Q}}
\newcommand{\tc}{{\tilde c}}
\newcommand{\zetab}{{\bar \zeta}}
\newcommand{\zb}{{\bar z}}
\newcommand{\qb}{{\bar Q}}
\begin{document}
\begin{titlepage}
\renewcommand{\thefootnote}{\fnsymbol{footnote}}
\begin{center}
{\headfontb Q-Exact Actions for BF Theories}\footnote{This
work is supported in
part by funds
provided by the
U. S. Department of Energy (D.O.E.) under contract
\#DE-AC02-76ER03069.}

\end{center}
\vskip 0.3truein
\begin{center}
{
{\Large R}{OGER}
		    {\Large B}{ROOKS}
{ AND}
{\Large C}{LAUDIO}
		   {\Large L}{UCCHESI}\footnote{Supported by the
Swiss
National Science Foundation.}
}
\end{center}
\begin{center}
{\addfont{Center for Theoretical Physics,}}\\
{\addfont{Laboratory for Nuclear Science}}\\
{\addfont{and Department of Physics,}}\\
{\addfont{Massachusetts Institute of Technology}}\\
{\addfont{Cambridge, Massachusetts 02139 U.S.A.}}
\end{center}
\vskip 0.5truein
\begin{abstract}
The actions for all classical (and consequently quantum)
$BF$  theories on
$n$-manifolds is proven to be given by anti-commutators of
hermitian,
nilpotent, scalar fermionic charges with Grassmann-odd
functionals.  In
order to show this, the space of fields in the theory must be
enlarged
to include ``mass terms'' for new, non-dynamical,
Grassmann-odd fields.
The implications of this result on  observables are examined.
\vskip 0.5truein
\leftline{CTP \# 2271  \hfill December 1993}
\smallskip
\leftline{hep-th/9401005}
\end{abstract}
\end{titlepage}
\setcounter{footnote}{0}

\section{Introduction}
\label{sec1}
\forcepar
Topological field theories (TFT's) are usually classified
as belonging to either one of two distinct classes.
One such class  called ``topological quantum
field theories" (TQFT's)\footnote{TQFT's are sometimes
referred to as
Donaldson-Witten or
cohomological field theories.}, is characterized by the fact
that the
actions are expressible as the anti-commutator of a
cohomology charge, $Q$,
with some field functional \cite{Wit} (we will refer to such
actions as being
$Q$-exact). Furthermore,
the (formal) argument establishing the topological invariance
of
correlation functions of observables (for a review see ref.
\cite{BBRT}) is based on the existence of the cohomology
operator.
The other class of TFT's is comprised of the $BF$ theories.
Until our work it
was not understood how to express the entire quantum
$BF$ action  as being
$Q$-exact.
However, we recently showed that two and three dimensional, abelian $BF$
theories and TQFT's
are related via a map
that inverts the Grassmann parity of the fields \cite{us}. This
yields a
unified picture for the two classes of TFT's outlined
above. Hence, the cohomology structure present in TQFT's
should
also exist in  $BF$ theories and it is an appealing
task
to derive it. In the present article, as part of establishing the
$Q$-exactness of
all $BF$ theories, we give the explicit calculus for such a
cohomology charge. We also show how this result is
intertwined with the topological invariance\footnote{Let us
recall that the $BF$ action, as well as the correlation
functions of the (Wilson loop) observables for the fields $B$
and $\ca$ [$F$ being the (covariant) exterior derivative of
$\ca$], are
topological \cite{HS}. In fact, correlation functions of
observables compute the intersection numbers of cycles on
the space-time manifold.} of the theory.

Other (partial) results are known which indicate that such a
cohomology should exist. In the two dimensional $BF$
theory
an equivariant cohomology operator was found \cite{Wit1}
by
enlarging the space of fields to include a Grassmann-odd
$1$-form.
However, in this extension and
its higher dimensional analog \cite{me}, position
independence could not be proven
for the correlation functions of operators which depend only
on the $\ca$ field. Furthermore, the classical action cannot
be written as
an anti-commutator with a nilpotent charge.  Contrarily, the
cohomology
that we present in this paper for the $BF$ theory has, as its
elements,
observables constructed most generally from the
$k$-cycle integrals over
the $\ca$ (and $B$) field(s), their exterior derivatives and
wedge products.

As we are using TQFT's as a paradigm,
the cohomology operator we seek, call it $Q$, must be a
scalar, hermitian, nilpotent symmetry generator.
In addition, the (covariant) exterior derivative of $\ca$ must
be $Q$-exact\footnote{For example, in the two dimensional
case, where $\ca=A$ is a
gauge field, $F$
must be the (anti)\-com\-mutator of  $Q$ with some
field.}.

With the above as a background, we apply our results
about
$BF$ theories also to actions
which are given by the square of  $F$ and can be written
as $BF$ actions by integrating out the $B$ field. For
example, the action for the two-dimensional Yang-Mills
theory (${\rm YM}_2$)
can be written in this sense as $S=\int_M(B\wedge F -\half B^2)$.

This paper is structured as follows. In Section \ref{sec1b}, we shall first
develop  our ansatz based upon TQFT's of constant
maps from Riemann surfaces to $\IR^d$.  Next, in Section
\ref{sec2}, these
observations are used to motivate our derivation of a
$Q$-exact action
for the two dimensional non-abelian $BF$ theory.  Then, in
Section \ref{sec3},
we  generalize the construction
to higher dimensional theories and analyze consequences on the observables.  In
Section \ref{sec4},
we investigate the twisted and untwisted supersymmetries in
the
two-dimensional  case.  We conclude in Section \ref{sec5}.
Throughout this
work our notation is as follows:
\medskip
\begin{center}
\begin{tabular}{|l|l|}\hline
OBJECT&{\hskip 2cm}DEFINITION\\ \hline\hline
$(\delta)~d$&(adjoint) exterior derivative\\ \hline
$d_A$&covariant exterior derivative\\ \hline
$\dual$&Hodge dual operator: $\dual{\hskip
-11pt}\dual~\equiv~ (-)^{k(n-k)}$\\
\hline
$\alpha_0$&parametrization from $BF$ to $F^2$ theories\\
\hline
$\Phi_{(k)}$&a generic form of degree $k$.\\ \hline
\end{tabular}
\end{center}
\centerline{Table 1: Notation}
\vskip 0.5truein
\setcounter{equation}{0}
\section{A Hint From TQFT's of Maps}\label{sec1b}
\forcepar
Before starting the investigation of the gauge theory case,
let us begin
by recalling the construction of topological sigma models
(see for instance \cite{BBRT}).  We perform this  exercise to
motivate the
transformations we shall be using in the
sections ahead.  The peculiarity here is
that we gauge fix to a constant the map $X^I$ ($I=1,\ldots,d$) from a
Riemann surface
$\Sigma$ to $\IR^d$. That is, we gauge fix the local
symmetry
$X^I\to X^I + \epsilon_X^I$  via introducing
the action
\be
S^{\rm map}_{0}\seq \{Q^{\rm map}_0,\int_\Sigma
\rho^I\wedge\dual (d X^I \mi
\half b^I)\}\ \ .
\ee
It turns out that this action is further invariant under the
local symmetry
$\rho^I\to\rho^I + \dual d\epsilon_\rho^I$, which we
gauge-fix by
imposing $d\rho^I=0$. The resulting action is
\bea
S^{\rm map} &\seq&\{Q^{\rm map},\int_\Sigma
(\rho^I\wedge\dual (dX^I \mi \half
b^I) \pls \varphi^I(d\rho^I\pls\half \alpha_0 \beta^I)~)\}\nen
&\seq& \int_\Sigma \biggl(\half dX^I\wedge \dual d X^I \pls
\half
d\varphi^I\wedge\dual d\varphi^I
 \pls \beta^I (d\rho^I \pls \half \alpha_0
\beta^I)\nen
&&\qquad \mi \rho^I\wedge \dual
d\lambda^I\pls {\varphi}^I
d\dual d{\tilde\varphi^I}\biggr)\ \ ,
\eea
where the BRST charge $Q^{\rm map}$ acts as
\be[Q^{\rm map},X^I]\seq \lambda^I\ \ ,\quad
[Q^{\rm map},\varphi^I]\seq\beta^I\ \ ,\quad
\{Q^{\rm map},\rho^I\}\seq b^I +\dual d{\tilde\varphi}^I\ \
.\label{Qmap}
\ee
In the final expression  for $S^{\rm map}$ we have
integrated out the auxiliary
field $b^I$ yielding $b^I=dX^I$ and discarded a surface term.
Notice that the $\beta^2$-term vanishes
due to the Grassmann-odd nature of $\beta$; we have left it
in for comparison
with later expressions.
The field content of the action $S^{\rm map}$
is summarized in the following Table:
\medskip
\begin{center}
\begin{tabular}{|c|c|c|c|}\hline
    &FORM&GRASSMANN\\
FIELD&DEGREE&PARITY\\ \hline\hline
$\rho$          &$1$&odd  \\ \hline
$\beta$         &$0$&odd  \\ \hline
$\lambda$       &$0$&odd  \\ \hline
$\varphi$       &$0$&even \\ \hline
$\tilde\varphi$ &$0$&even \\ \hline
$X$          &$0$&even \\ \hline
$b$             &$1$&even \\ \hline
\end{tabular}
\end{center}
\centerline{Table 2: Form degrees and Grassmann parities of
the fields in
$S^{\rm map}$.}
\medskip

Let us now consider this action
irrespective of its origin and reverse the Grassmann parity of
the fields.
The first two terms in $S^{\rm map}$ vanish. Next we rename the fields as
$(\rho,\beta,\lambda,\varphi,{\tilde \varphi},X,b)\to
(A,B,\Lambda,c,\tc,\chi,\psi)$.  Having done this,
we recognize (for the case $d=1$) the Maxwell (or abelian
$BF$) action in the
Landau gauge.
Hence, for the non-abelian case, this suggests that in order
to obtain the gauge-fixed $BF$ action as a $Q^{\rm
map}$-exact
expression, we should enlarge the space of fields to include
the Grassmann-odd counterpart of $X$.
\vskip 0.5truein
\setcounter{equation}{0}
\section{$Q$-Exact Two-dimensional $BF$ Theory}
\label{sec2}
\forcepar
In this section, we shall be concerned with the non-abelian theory
in two dimensions, for which we now formalize the observations
presented above.
Consider  a space of fields spanned by those appearing in the
gauge fixed $BF$/\ym theory and enlarge it to include two new
fields,
$\chi$ and $\eta$. These are Grassmann-odd and take value in the
adjoint representation of
the gauge group; they will turn out not to be dynamical. We
summarize our
space of fields in the following table:
\medskip
\begin{center}
\begin{tabular}{|c|c|c|c|}\hline
         &FORM&GRASSMANN\\
FIELD&DEGREE&PARITY\\ \hline\hline
$A$&$1$&even\\ \hline
$B$&$0$&even\\ \hline
$\Lambda$&$0$&even\\ \hline
$c$&$0$&odd\\ \hline
$\tilde c$&$0$&odd\\ \hline
$\chi$&$0$&odd\\ \hline
$\eta$&$2$&odd\\ \hline
\end{tabular}
\end{center}
\centerline{Table 3: Enlarged space of fields for $BF/{\rm YM}_2$.}
\medskip
\noindent On this space of fields we define a scalar fermionic
charge $\cq$ by:
\be
\{\cq ,\chi\}=\Lambda\ \ ,\qquad \{\cq ,c\}=B\ \ ,\qquad
[\cq ,A]=\dual d\tc\ \ ,\label{TRANS}
\ee
and $[\cq,(\eta,\Lambda,B,\tc)\}= 0$, as motivated by the
transformations (\ref{Qmap}).
Nilpotency is immediate: $\cq^2=0$.  Furthermore, we take all the
fields to be hermitian so that
$\cq^\dagger=\cq$.  The action of $\cq$ on the fields suggests that it
is related to the abelian anti-BRST symmetry of the gauge-fixed $BF$ theory
\cite{us,BFbrst}.  Using this cohomological structure, we define, in
analogy with $S^{\rm map}$ of Section \ref{sec1b}, the
$\cq$-exact action:
\be
S(\alpha_0)\seq \{\cq,\int_\Sigma Tr\,(\,c\,(F\pls \half\alpha_0
\dual
B)\pls \chi\dual\delta A)\}\ \ ,\label{SGEN}
\ee
where $\alpha_0$ is an arbitrary parameter.  Employing the
transformation rules (\ref{TRANS}) we find, up to surface terms
involving the fields $c$ and $\tc$:
\be
S(\alpha_0)\seq \int_\Sigma Tr(B F \pls \half\alpha_0 B^2\pls
\Lambda \dual\delta A
\mi \tc \dual\delta d_A c)\ \ ,\label{SALPHA}
\ee
where $d_A$ is the
covariant exterior derivative.  Notice that $\chi$ does not
appear in the final form of the action, even in a purely auxiliary
capacity\footnote{Indeed, $\chi$ only appears at an intermediate
stage of the calculation in the vanishing
term $\chi\delta\dual d\tc= \chi\delta^2{}\dual \tc=0$.}. We
recognize
the first term as the usual pure $BF$ action,
the second is analogous to the $\beta^2$ term in $S^{\rm map}$, the third
implements the Landau gauge condition and the last
term is the corresponding Faddeev-Popov action.
The arbitrary parameter $\alpha_0$ interpolates between \ym and
$BF$
theories. Indeed, with $\alpha_0=2$ and after integrating out the
$B$-field, (\ref{SALPHA}) is just the gauge-fixed \ym
action,
whereas $\alpha_0=0$ yields the gauge-fixed $BF$ theory in two dimensions.

{}From (\ref{SGEN}) it would appear that the partition function of \ym is
metric independent following the usual arguments from TQFT's.  We know that
this is not the case, however.
Hence, it must be that the symmetry generated by
$\cq$ is broken by an anomaly or a non-trivial measure.  This is the
case,
indeed, and as we will now see.

{}From (\ref{TRANS}), it is natural to  grade the space of
fields by introducing a $U(1)$ charge akin to ghost number, and
assigned as follows:
$(A,B,\Lambda,c,\tc,\chi,\eta) \leftrightarrow (0,0,0,1,-1,1,-1) $;
$\cq$ itself carries charge $-1$. Since $\chi$ has
charge $1$, a partner $\eta$ must be introduced in order to
form a
$U(1)$ invariant functional measure.  Although $\chi$ appears in
(\ref{SGEN}), neither field appears in the final form of the action,
(\ref{SALPHA}). However, as they were
needed in establishing the $\cq$-exactness of the action, we
must integrate over them in the partition function. And since these fields
are fermionic, in order for
the partition function not to be zero, we must introduce a
non-trivial measure. To wit, we construct the measure as
\be
\pint_{\chi,\eta}\seqv \pint [\fd\chi][\fd \eta]
\exp{\biggl(\int_\Sigma Tr (\eta \chi)\biggr)}\ \ .\label{MESURE}
\ee
This measure explicitly breaks the $\cq$ symmetry.
We now use it to define
the partition function on the enlarged space of fields (see Table 3) by
simply adding the exponent in the measure (\ref{MESURE})
to the action. This yields:
\be
Z\seq \pint [\fd A][\fd B][\fd \Lambda][\fd c][\fd \tc][\fd
\chi][\fd \eta] \exp{\biggl(\mi(S_0(\alpha_0)+S_{GF})\biggr)}\ \ ,
\ee
where
\bea
S_0(\alpha_0)&\seq& \int_\Sigma Tr\,(\eta\chi\pls B F \pls
\half\alpha_0
B^2)\ \ ,\nen
S_{GF}&\seq&\int_\Sigma Tr\,(\Lambda \dual\delta A \mi \tc
\dual\delta d_A c) \ \
.\label{Z}
\eea
$S_0(\alpha_0)$ is interpreted as the gauge
invariant classical action
and $S_{GF}$ arises from gauge-fixing the Yang-Mills
symmetry.

The classical action $S_0(\alpha_0)$ is
invariant under new sets of Grassmann-odd symmetries generated by $Q$ and
$\tq$:
\bea
[Q,B]&= &-\chi\ \ ,\qquad \{ Q,\eta\} =
F \pls \alpha_0 \dual B\ \ ,\nennn
[\tq,B]&= &\dual\eta\ \ ,\qquad\, \{ \tq,\chi\} =
\dual F\pls\alpha_0 B\ \ .
\label{Qtrans}
\eea
(Notice that there is no factor $\half$ multiplying $\alpha_0$).
Heuristically, $\tq$ is dual to $Q$.
Expressing $S_0(\alpha_0)$ in terms of these charges
yields the following interesting results:
\bea
S_0(\alpha_0)&\seq&\{Q,\int_\Sigma Tr\, (B\eta) \} \mi \half \,
\alpha_0 \int_\Sigma Tr\, (B^2) \nen
&\seq&\{\tq,\int_\Sigma Tr\, (B\chi) \} \mi \half \,\alpha_0
\int_\Sigma Tr\, (B^2 )\ \ ,
\label{Sexact}
\eea
from which we immediately learn two things.  First,  the classical $BF$
action on the enlarged space of fields
$S_0(\alpha_0=0)$ is $Q$- and/or
$\tq$-exact.  Secondly,  as expected, the \ym action
($\alpha_0\neq0$) is neither $Q$- nor $\tq$-exact.
This is consistent with the observation that the transformations
$Q$ and $\tq$ in (\ref{Qtrans}) are nilpotent only when
$\alpha_0=0$, that
is only in the case of the $BF$ theory. The physical states of the
$BF$ theory are hence elements of the $Q$-cohomology (and/or of the
$\tq$-cohomology), unlike the physical states of ${\rm YM}_2$.

\vskip 0.5truein
\setcounter{equation}{0}
\section{Higher Dimensions and Observables}
\label{sec3}
\forcepar
Generalization of the results presented above to higher dimensions is
immediate.  Given a
$k$-form $\ca$ which transforms homogeneously in the adjoint
representation
of a gauge group\footnote{Of course, we can also take $\ca$ and $B$ not to take
values in a gauge group, in which case, $F=d\ca$.}, and such that $F\equiv
d_A\ca$ is gauge covariant
($A$ is a background gauge field), the analog of
$S_0(\alpha_0)$ on an $n$-manifold $M$ is
\be
S_0^{(n)}({\alpha_0})\!=\!\{Q,\!\int_M
\!Tr\, (B_{(n-k-1)}\wedge\eta_{(k+1)})\}
-\half \alpha_0 \!\int_M \!Tr\, (B_{(n-k-1)}\wedge\!\dual
B_{(n-k-1)}) \ \ ,
\ee
and yields:
\bea
S_0^{(n)}({\alpha_0})&\seq&\int_M Tr\,\biggl(
\eta_{(k+1)}\wedge \chi_{(n-k-1)}
\pls B_{(n-k-1)}\wedge F_{(k+1)}\nen
&&\qquad\quad\pls\half\,\alpha_0 \,B_{(n-k-1)}
\wedge\dual B_{(n-k-1)}\biggr)\ \ ,
\eea
where the generator $Q$ acts as
\bea
[Q,B_{(n-k-1)}]&\seq& -(-)^{(n-k-1)(k-1)}\chi_{(n-k-1)}\ \ ,
\nennn
\{Q,\eta_{(k+1)}\}&\seq& \cf_{(k+1)} \pls \alpha_0
\dual B_{(n-k-1)}\ \ ,
\label{QnTrans}
\eea
and we omit writing the corresponding transformations generated by
$\tq$.  Based on the $Q$-exactness of the $BF$ action, we conclude that its
partition function is independent of the background gauge field, $A$.

Using the results above, we now address the question of the
observables
in a $BF$ theory on an $n$-manifold. Let us start with $\co_\ca$, by which we
denote gauge invariant
operators constructed out of invariant polynomial  functions
$\cp(F)$ of $F$ and  generalized Wilson
loop operators, $W_\Gamma(\ca)$ (where
$\Gamma$ is a representative homology $k$-cycle). An example of such an
operator
$\co_\ca$
is $ \cp(F)\otimes W_{\Gamma}(\ca)$.
Those $\co_\ca$ for which $\cp$ is at least of degree one,
will have vanishing
correlation functions amongst themselves.  This is due to the
fact that $F$
is $Q$-exact and all functions of $\ca$ are $Q$-closed.  This
is not true,
however, for observables composed of the gauge invariant
polynomials of those
$\ca$'s for which $F=0$; for example, flat connections in
the case $k\neq 1$.
Derivatives of the latter with respect to the collective
coordinates, on which
they may depend, can be shown to form gauge invariant
polynomial observables
(see ref. \cite{Gilme}).  Our analysis does not preclude
these from being
non-trivial.

Restricting ourselves to the Wilson loop content of $\co_\ca$, we find
that
correlation
functions of the observables $W_{\Gamma}(\ca)$ can only depend on
the
homotopy class of the $\Gamma$'s. The standard proof of this
statement
\cite{BT1} makes
explicit use of the restriction of the path integral to $F=0$
solutions
(arising from the integration over $B$)\footnote{Of course,
this assumes that
apart from gauge fixing terms, there is no additional
$B$-dependence in the
path integral under investigation.}.  Given that under a homotopically
trivial  perturbation in $\Gamma$, the change in $W_\Gamma(\ca)$
depends\footnote{This can be found
in equation (2.8) of ref. \cite{BT1}.} on
$F$, we conclude that such
a change will be $Q$-exact.  Consequently, its effect in a
correlation function
will vanish.  Thus we have proven the statement about the
dependence of
Wilson loops on the homotopy class without appealing to the
``on-shell'' $F=0$ condition.  Proceeding further, we can
readily see that
the $W_\Gamma(\ca)$ can only depend on the $\ca$
zero-modes (solutions of
$F=0$).  This is due to the fact that since these do not
contribute to the action, they do not appear in the transformations
(\ref{QnTrans}).  In turn, this means
that they are automatically in the $Q$-cohomology.

Next, we restrict our attention to the subset of
observables $\co_B$, by which we denote operators which depend,
in a gauge invariant manner, on $B$.
We identify the zero-modes of $B$ as the solutions of $d_AB=0$.
The argument is identical to the above: since these
do not contribute to the action, they are inert under the
action of $Q$.
This means that only gauge invariant functions of the $B$
zero-modes will be in the
$Q$-cohomology.

Having studied separately the cases of  $\co_\ca$ and
$\co_B$, we now turn to observables which depend on both $\ca$ and $B$.  Since
all observables are in the $Q$-cohomology, our
previous discussions regarding the $\co_\ca$ case applies also in the
mixed
case (remember that we did not have to appeal to the ``on-shell"
$F=0$ or $d_AB=0$ conditions).
As a consequence, the observables of a $BF$ theory on an $n$-manifold
can only depend on the zero-modes of $\ca$ and
$B$ and on the homotopy classes of cycles over which they are
integrated.

\vskip 0.5truein
\setcounter{equation}{0}
\section{Twisting in Two Dimensions}
\label{sec4}

Let us now return to the discussion of two dimensional space-times. The
symmetries (\ref{QnTrans}) hold on an arbitrary Riemann
surface $\Sigma$. If the manifold is
flat, however, we find that the action $S_0(\alpha_0)$
(\ref{Sexact}) is also invariant  under
\be
\{Q_{(1)},\dual\eta\}\seq d_A B\ \ ,\qquad [Q_{(1)},A]=\dual\chi\ \ ,
\ee
where $Q_{(1)}$ carries form-degree one and is not nilpotent.
Of greater value is the fact that with $\alpha_0=0$,
we have the algebra:
\be
\{\tq,Q_{(1)}\}\seq d_A\ \ ,\qquad \{Q,Q_{(1)}\}\seq 0\ \ .
\ee
Due to the vanishing of the anticommutator $\{Q,Q_{(1)}\}$,
this is not a twisted
$N=2$ supersymmetry algebra. That this is indeed not the case may also be
seen from
the field content of the action $S_0$, and it is precisely by taking
new
fields into account that we shall obtain the twisted
supersymmetric action.  In ref.
\cite{Wit1}, an anti-commuting $1$-form, $\psi$, was
introduced in order to
define the symplectic form on the space  of gauge
fields.  This field
transforms in the adjoint representation of the gauge
group and forms a
basis for $1$-forms on that space.  Let us  add its action to
$S_0$, and define our new action to be (on arbitrary Riemann surfaces)
\be
S_0'(\alpha_0)\seq \int_\Sigma Tr\,\biggl(
\eta\chi\pls B F \pls \half\alpha_0 B^2 \pls\half
\psi\wedge\psi\biggr)\ \ ,
\label{Spsi}
\ee
with four, real anti-commuting degrees of
freedom. In addition to being invariant under the
transformations (\ref{Qtrans}), the action (\ref{Spsi}) possesses
another set of symmetries \cite{Wit1} which do not
affect the $\eta$ and $\chi$ fields:
\be
\begin{array}{rclcrcl}
[Q_\psi,A]\seqq\psi\ \ , \qq\{Q_\psi,\psi\}\seqq
d_A B\ \ ,\nen
[\tq_\psi,A]\seqq\dual\psi\ \ , \qq\{\tq_\psi,\psi\}\seqq
\dual d_A B\ \ .
\end{array}
\ee

The action $S_0'(\alpha_0=0)$ is actually the twisted form of
the supersymmetric action
\be
S_{\rm susy}\seq \frac{i}{2}\int d^{2}\!z \,Tr\,
\biggl(B F_{z\zb}\pls \zeta_+ \zetab_- \pls
\zeta_-\zetab_+\biggr)\ \ ,\label{susyaction}
\ee
where $\zeta_\pm$ are the Weyl components of a complex
spin-$\half$ field. Let us now prove this statement. We start from the fact
that $S_{\rm susy}$ is invariant under the $N=2$ supersymmetry
transformations:
\be
\begin{array}{rclcrcl}
[Q_+,A_\zb]\seqq \zeta_-\ \ ,
  \qq [\qb_+,A_z]\seqq -\zetab_-\ \ ,\nenn
[Q_+,B]\seqq \zeta_+\ \ ,
  \qq [\qb_+,B]\seqq \zetab_+\ \ ,\nenn
\{Q_+,\zetab_-\}\seqq F_{z\zb}\ \ ,
  \qq \{\qb_+,\zeta_-\}\seqq F_{z\zb}\ \ ,\nenn
\{Q_+,\zetab_+\}\seqq D_z B\ \ ,
  \qq \{\qb_+,\zeta_+\}\seqq D_\zb B\ \ ,\nenn
[Q_-,A_z]\seqq \zeta_+\ \ ,
  \qq [\qb_-,A_z]\seqq -\zetab_+\ \ ,\nenn
[Q_-,B]\seqq \zeta_-\ \ ,
  \qq [\qb_-,B]\seqq \zetab_-\ \ ,\nenn
\{Q_-,\zetab_-\}\seqq - D_\zb B\ \ ,
  \qq \{\qb_-,\zeta_-\}\seqq - D_z B\ \ ,\nenn
\{Q_-,\zetab_+\}\seqq F_{z\zb}\ \ ,
  \qq \{\qb_-,\zeta_+\}\seqq -F_{z\zb}\ \ ,
\end{array}
\ee
where $D_z$ is the gauge covariant derivative.
In addition to being Lorentz invariant ($[M,\zeta_\pm]=\half
\zeta_\pm$,
where $M$ is the Lorentz generator), the action (\ref{susyaction}) is
also symmetric under a fermion number $J$ for which the
assignments
$(\zeta_+,\zetab_+,\zeta_-,\zetab_-)
\leftrightarrow (1,-1,1,-1)$ are made. Twisting these fields
by re-defining the Lorentz generator to be $M'\equiv M-\half J$,
we find that
\be
[M',(\zeta_+,\zetab_+,\zeta_-,\zetab_-)]=(0,\zetab_+,-\zeta_-,
0)\ \ .
\ee
That is, $\zeta_+$ transforms as the $z$-component of a vector
whereas $\zeta_-$ behaves
as its $\zb$-component.  Thus we rename
\be
(\zeta_+,\zetab_+,\zeta_-,\zetab_-) \longrightarrow
(\chi,-\inv{\sqrt{2}}\psi_\zb,-\inv{\sqrt{2}} \psi_z,-\eta_{z\zb})\ \ ,
\ee
and the action becomes
\be
S_{\rm susy}^{\rm twisted} \seq
\frac{i}{2} \int d^2\!z\, Tr\,\biggl( \eta_{z\zb}\chi \pls B
F_{z\zb}\pls\half \psi_z\psi_\zb\biggl)\seq S'_0(\alpha_0=0) \ \ .
\ee
This turns out to be the flat space-time version of $S_0'(\alpha_0=0)$ from
(\ref{Spsi}), thus proving the above statement.

We now comment on the nature of the supersymmetries in the last
two, untwisted and twisted, actions.  First, $S_0'(\alpha_0=0)$
cannot be written as a $(Q+Q_\psi)$-exact
expression [even though, as we have shown,
$S_0(\alpha_0=0)$ can be written as a
$Q$-exact expression]. This seems to be in
contradiction with the fact that $S_0'(\alpha_0=0)$ is the
twisted version of  $S_{susy}$. Actually, it is possible to show
that $S_{susy}$ is itself given by the action of a spin-$\half$
generator on some  functional. This is most simply done in $N=1$
superspace with two matter supermultiplets,
the first one being the Yang-Mills supermultiplet and the
second one being a spinor supermultiplet.
The YM supermultiplet is given by the spinor superfield strength,
$W_\alpha$ (greek letters are used as spinor indices)
in which  the lowest component is the
gaugino $\lambda_\alpha$ and the middle component is the
bosonic field strength, $F$. The spinor superfield, $\cb^\alpha$, has
components
$(\kappa^\alpha,B, \beta^\alpha)$.  The field $\zeta_\alpha$
which appears in
$S_{susy}$ is the complex combination of $\beta_\alpha$ (the top component of
$\cb_\alpha$)
and $\lambda_\alpha$.
$\kappa^\alpha$ appears in the component form of the
superspace lagrangian in a ``kinetic'' term for the gaugino
$\lambda_\alpha$, as $i\kappa^\alpha\,{\slash{\hskip
-0.27cm}D}_{\alpha\beta}\lambda^\beta$.
This term is missing in $S_{susy}$.  Likewise, its twisted
version is
missing in $S_0'(\alpha_0=0)$.  Had we added such a term to
the latter action, we would have obtained the twisted
version of the $N=2$ supersymmetric $BF$ action,  which is
a full fledged Donaldson-Witten theory. The lack of this term is the
reason that $S_0'(\alpha_0=0)$ is not $(Q +Q_\psi)$-exact.
Nevertheless, we reiterate that $S_0(\alpha_0=0)$ is $Q$-exact.
\vskip 0.5truein
\section{Conclusions}
\label{sec5}
\forcepar
We have proven that all $BF$ theories on
$n$-dimensional manifolds
can be written as the anti-commutator of a Grassmann-odd,
hermitian, nilpotent
charge with some functional.  This result was obtained by enlarging
the space of fields to include two additional Grassmann-odd fields.
The cohomology structure
allows us to recover the known restrictions on the observables, namely that
they may depend only
on zero-modes of the fields and the homotopy classes of
the homology cycles used in constructing them.
Since our parametrization of the actions also includes the $F^2$-theories, for
example Yang-Mills, it immediately follows that the latter theories are not
$Q$-exact.

The $BF$ theory in the Landau gauge is  perturbatively finite in dimensions
less than six \cite{BFbrst,BMag,MagS,GMS}. The two-dimensional case, treated in
\cite{BMag}, deserves special attention due to its IR singularities, but the
proof of finiteness can nevertheless be maintained. Since the Grassmann-odd
fields we added to the $BF$ action in order to prove its $Q$-exactness are
non-propagating and uncoupled to the original fields, they are excluded from
loop processes.
Therefore, our result establishes that the $BF$ action is also perturbatively
$Q$-exact.

\end{document}